\documentclass[12pt,preprint]{aastex}
\def \ev{~\rm{eV}}

\def \cm{~\rm{cm}}
\def \s{~\rm{s}}
\def \km{~\rm{km}}

\def \K{~\rm{K}}
\def \g{~\rm{g}}

\def \AU{~\rm{AU}}
\def \erg{~\rm{erg}}

\def \yr{~\rm{yr}}

\begin{document}

\title{ACCRETION ONTO THE COMPANION OF ETA CARINAE DURING THE SPECTROSCOPIC EVENT. IV.
\newline
THE DISAPPEARANCE OF HIGHLY IONIZED LINES}

\author{Noam Soker\altaffilmark{1}}

\altaffiltext{1}{Department of Physics, Technion$-$Israel
Institute of Technology, Haifa 32000 Israel;
soker@physics.technion.ac.il.}

\begin{abstract}

We show that the rapid and large decrease in the intensity of high-ionization
emission lines from the $\eta$ Carinae massive binary system can be explained
by the accretion model.
These emission lines are emitted by material in the nebula around the binary system
that is being ionized by radiation from the hot secondary star.
The emission lines suffer three months long deep fading every 5.54 year, assumed
to be the orbital period of the binary system.
In the accretion model,  for $\sim 70$~day the less massive secondary star is accreting
mass from the primary wind instead of blowing its fast wind.
The accretion event has two effects that substantially reduce the
high-energy ionizing radiation flux from the secondary star.
(1) The accreted mass absorbs a larger fraction of the ionizing flux.
(2) The accreted mass forms a temporarily blanked around the secondary star
that increases its effective radius, hence lowering its effective temperature
and the flux of high energy photons.
This explanation is compatible with the
fading of the emission lines at the same time the X-ray is
declining to its minimum, and with the fading being less pronounced in
the polar directions.
\end{abstract}

% ==========================================================
\section{INTRODUCTION}
\label{sec:intro}
% ==========================================================

The two winds blown by the massive stellar binary system $\eta$ Car are
major players in the $5.54 \yr$ light periodicity.
The periodic variation is observed from the IR (e.g., Whitelock et al.\ 2004)
to the X-ray band (Corcoran 2005; Corcoran et al.\ 2001, 2004a,b).
According to most models, the periodic winds interaction behavior follows
the 5.54~years periodic change in the orbital separation in this highly eccentric,
$e \simeq 0.9$, binary system (e.g., Hillier et al. 2006).
The X-ray deep minimum lasts $\sim 70$~day and occurs more or less simultaneously
with the spectroscopic event (e.g., Damineli et al.\ 2000),
defined by the fading, or even disappearance, of high-ionization emission lines  (e.g.,
Fe~III~$\lambda$1895, Fe~III~$\lambda$4701,
Ne~III~$\lambda$1747-54, Si~III~$\lambda$1892, Zanella et al.
1984; He~I~$\lambda$10830, Damineli 1996; He~I~$\lambda$6678,
Damineli et al. 2000; and many more lines listed by Damineli et al. 1998).
The spectroscopic event includes changes in the continuum and other lines
(e.g., Martin et al. 2006,a,b; Davidson et a. 2005).
The X-ray minimum and spectroscopic event are assumed to occur near
periastron passages.

The change in the orbital separation and absorbtion by the wind
or a presumed eclipse by one or two of the stars cannot account for the
$\sim 70$~day long X-ray minimum (Ishibashi et al.\ 2003; Hamaguchi et al.\ 2005).
To account for the deep X-ray minimum Soker (2005a,b) has suggested that for
$\sim 10$ weeks near periastron passages the secondary does accrete
mass from the primary wind.
The collision region of the two winds, a very fast wind from the secondary star and
a slower wind from the primary star, is responsible for the X-ray emission along
most of the orbit
(Corcoran et al.\ 2001; and Pittard \& Corcoran 2002;Akashi et al.\ 2006).
According to the accretion model the deep minimum is assumed to result
from the collapse of the collision region of the two winds onto
the secondary star. This process is assumed to shut down the
secondary wind, hence the main X-ray source.
Akashi et al. (2006) showed that this assumption provides a
phenomenological description of the X-ray behavior around the minimum.
The accretion model was applied also (Soker \& Behar 2006) to explain the appearance
of the He~II~$\lambda$4686 emission line before the event and its disappearance
during the event (Steiner \& Damineli 2004; Martin et al. 2006b).

{{{  One of the motivations for developing the accretion model for
the spectroscopic event is the {{{ problem of the single star
shell-ejection model }}}  (Zanella et al. 1984) to account for
the X-ray decline (Akashi et al. 2006). Hamaguchi et al. (2005)
observed $\eta$ Car 24 days into the minimum (on 2003-07-22) with
XMM-Newton. Despite the huge decline, the X-ray spectrum does not
become harder, as would be expected if absorption is responsible
for the decline in X-ray intensity. This shows that absorption
cannot be the main reason for the X-ray decline. After another
several days the X-ray luminosity starts to increase for a while
(Corcoran 2005), and becomes harder (Corcoran 2005; Hamaguchi et
al. 2005). This again contradicts an absorption effect in which a
rise in flux (less absorption) would be accompanied by softening.
In addition, in the shell ejection model the stellar primary mass
loss rate should increase by a factor of $\sim 20$ (Corcoran et
al. 2001). There is a larger question regarding the mechanism
capable of increasing the mass loss rate by this large factor
(Soker 2005a). In any case, an increase by a factor of $\sim 20$
in the primary mass loss rate in the equatorial plane will make
accretion onto the secondary inevitable. Even without this
increase accretion is suggested to occur in the accretion model.
Therefore, any model for a periastron increase in primary mass
loss rate must consider accretion onto the companion. }}}
{{{ In addition, in a recent paper Nielsen et al. (2007) interpret their
spectroscopic data as strong support for binarity over a single
star shell-ejection. }}}

The accretion process {{{  might }}} lead to the formation of an accretion disk for a very
short time (Soker 2003, 2005a), and {{{  might }}} even lead to a transient launching of
two opposite jets,
{{{  as suggested theoretically (Soker 2005a; Akashi et al. 2006), and might have been
observed (Behar et al. 2006).  Basically, in a steady state Bondi-Hoyle type accretion
flow (see below) the accreted mass has not enough angular momentum to form an accretion disk.
However, stochastically accreted blobs at the onset of the accretion phase
might lead to the formation of a transient accretion disk, and possibly to
two jets.  }}}
Earlier suggestions for a disk in $\eta$ Car were made by van Genderen et al.\ (1994, 1999).
The accretion process proposed to occur for only $\sim 10$ weeks near periastron passage
in present $\eta$ Car is different from what was presumably happening during the Great Eruption.
The Great Eruption (Davidson \& Humphreys 1997) is the 20 years event that led
to the formation of the bipolar nebulae
around $\eta$ Car (the Homunculus) starting one hundred and seventy yeas ago.
%% The similarity of the Homunculus morphology to the morphologies
%% of many planetary nebulae and symbiotic nebulae suggests
%% that during the Great Eruption the secondary star accreted mass from the primary along
%% the entire orbit, and blew two jets which shaped the Homunculus (Soker 2001, 2004, 2007).

{{{  Few words on the accretion process are in place here.
The accretion flow structure is of a compact object of mass $M_2$, the secondary star,
moving with a speed $v_1$ through a cloud, the primary wind.
This case was studied by Bondi \& Hoyle (1944).
The gas flows toward the compact object and pass through a shock wave.
When the gas radiative cooling time is very short, as is the case here,
the shock wave bends backward, and most of the mass is accreted from a region
behind the star, called the accretion column (see fig. \ref{fwinds}).
For the present model, this implies that during the accretion process the
secondary ionizing radiation will also be blocked to directions opposite the primary direction.
The gas that is accreted is the gas that flows within a distance
of $< R_{acc2}$ from the accreting object.
Here $R_{\rm acc2}$ is the accretion radius, which is more or less the
Bondi-Hoyle accretion radius $R_{\rm acc2} \simeq 2 G M_2/v_1^2$.

For the typical parameters of $\eta$ Car, the accretion radius is always much smaller that
the orbital separation, $R_{\rm acc2} \la 0.2 r$ (Soker 2005b; Akashi et al. 2006).
Therefore, the latitude dependance of the primary mass loss rate and velocity
(Smith et al. 2003) is not significant for the model.
What is important is the primary wind portion that interacts with the secondary wind.
This is the portion that is blown near the equatorial plane, and its properties are
determined from the X-ray emission (Pittard \& Corcoran 2002; Akashi et al. 2006).

For the accretion process to take place, the primary wind should reach a distance
of $\la R_{\rm acc2}$ from the secondary. For most of the orbit the secondary wind
prevents the primary wind to reach this distance (Soker 2005b; Akashi et al. 2006).
However, near periastron passages the colliding region of the two winds gets
closer to the secondary star, and accretion is likely to take place (Soker 2005b).
A higher mass loss rate from the primary will push the colliding region
closer to the secondary, hence will make the accretion process more pronounced.

The initiation of the $\sim 10$ weeks accretion phase was discussed in Soker (2005b).
Until 3D numerical gas-dynamical simulations are performed, the initiation processes
of the accretion phase outlined below should be considered as a scenario, rather than
an established model.
Because of thermal instabilities dense large blobs are formed in the post-shock primary's wind
region near the stagnation point. As periastron is approached the colliding wind region near
the stagnation point gets closer to the secondary star, the secondary's gravity
influence on these blobs increases, and just prior to periastron
passage the blobs become bound to the secondary's, and fall onto it.
Very close, possibly $\sim 10$~day prior, to periastron passage the mass of the primary's wind
that is accreted is assumed to be large enough to shut down the secondary wind.
The assumed shut-down must be a non-linear process, because the mass accretion rate is smaller
than the mass loss rate of the secondary. As the secondary's wind no longer reaches the
previous stagnation region, the primary wind flows toward the secondary and
the Bondi-Hoyle type accretion flow stars.
A key assumption in the model is that blobs accreted near periastron passage shut down,
or substantially weaken, the secondary wind.
After $\sim 10$ weeks the orbital separation substantially increases such that the
mass accretion rate declines and the secondary wind rebuilds itself.
}}}

In the present paper we try to explain the fading and disappearance of
high-ionization emission lines, from the near UV to the near IR,
with the accretion model (Soker 2005b).
This is a 4th paper in a series of papers aiming at understanding the spectroscopic
event by an accretion process onto the secondary star during the event.
The reader should consult earlier papers for more details on the accretion process.

% ==========================================================
\section{THE BINARY SYSTEM}
\label{sec:binary}
% ==========================================================
The $\eta$ Car binary parameters used by us are as in the previous papers in this series
(Soker 2005b; Akashi et al.\ 2006; Soker \& Behar 2006). They are based on several papers
and taking into account the present disagreement on some
of the binary parameters (e.g., Ishibashi et al. 1999; Damineli et al. 2000;
Corcoran et al. 2001, 2004b; Hillier et al. 2001; Pittard \& Corcoran 2002;
Smith et al. 2004).
The assumed stellar masses are $M_1=120 M_\odot$, $M_2=30 M_\odot$, the eccentricity is
$e=0.9$, and orbital period 2024 days, hence the semi-major axis is
$a=16.64 \AU$, and the orbital separation at periastron is $r=1.66 \AU$.
The mass loss rates are $\dot M_1=3 \times 10^{-4} M_\odot \yr^{-1}$
and $\dot M_2 =10^{-5} M_\odot \yr^{-1}$.
The terminal wind speeds are taken to be $v_1=500 \km \s^{-1}$ and
$v_2=3000 \km \s^{-1}$.
{{{  It is assumed here, and in the previous papers in the series,
that the orbital plane is oriented in the same plane as the equatorial
zone of the primary's wind, and the equatorial plane of the Homunculus. }}}

The secondary can be assumed to be an O star.
Somewhat evolved main sequence O-stars with $M_2=30 M_\odot$
can have an effective temperature of $T_2 \simeq 40,000 \K$, and
a luminosity of $L_2 \simeq 3 \times 10^5 L_\odot$, hence a
radius of $R_2 \simeq 11 R_\odot$; such stars have mass loss rates of up
to $\sim 10^{-5} M_\odot \yr^{-1}$ (e.g., Repolust et al. 2004).
These estimates are associated with large uncertainties since most likely the
secondary underwent a massive accretion event during the Great Eruption
(Soker 2001, 2004, 2007), which ended $\sim 150$~yr ago and hence it is likely to be
out of thermal equilibrium.
Recently, Verner et al. (2005) deduced the following secondary properties:
$T_2 \simeq 37,200 \K$, $L_2 \simeq 9.3 \times 10^5 L_\odot$,
$R_2 \simeq 23.6 R_\odot$, $v_2=2000 \km \s^{-1}$, and
$\dot M_2 \simeq 8.5 \times 10^{-6} M_\odot \yr^{-1}$.
We follow Soker \& Behar (2006) and take
$R_2 = 20 R_\odot$ and $T_2=40,000 \K$.
The primary star is more luminous but it is larger, and its effective temperature
is much lower. Therefore, it is the secondary star that ionizes the gas that is
the source of the high-ionization emission lines (Verner et al. 2005).

To demonstrate the crucial role of the accretion of primary's wind by the secondary
star we examine the ionizing flux at two energies, corresponding to ionization
of hydrogen and helium.
The rate of ionizing photons per steradian, i.e., having energy of $h \nu > 13.6 \ev$
to ionize hydrogen and $h \nu > 24.6 \ev$ to ionize helium,
emitted by the secondary star depends on its temperature.
The dependance of the ratio of ionizing photons to stellar luminosity on effective
temperature (Schaerer \& de Koter 1997) can be fitted with a linear relation for
both helium and hydrogen ionizing photons in the range $35,000 \la T_2 \la 42,000 \K$.
The number of photons emitted per stellar energy output is given
by
\begin{eqnarray}
n(h\nu >13.6 \ev)= \left( 43   \frac{T_2}{40,000 \K} -30.4 \right) \times 10^9 ~
{\rm photon}\erg^{-1} \qquad 30,000 \la T_2 \la 42,000 \nonumber
\\
n(h\nu >24.6 \ev)= \left( 23 \frac{T_2}{40,000 \K} -20  \right)\times 10^9 ~
{\rm photon}\erg^{-1} \qquad 35,500 \la T_2 \la 42,000
\label{nerg}
\end{eqnarray}
We here give the ionizing photon rate at two temperatures which we consider
the bounds of reasonable values for the secondary in $\eta$ Car; a linear fit can be done
for any effective temperature in the above range.
\begin{equation}
 \dot N_{i2} =\frac {\dot N_{i2-t}}{4 \pi} =
 \frac{L_2}{9\times 10^{5} L_\odot} \s^{-1} {\rm sr}^{-1}\left\{
\begin{array}{ll}
   3.5 \times 10^{48}    & \quad  {\rm Hydrogen} \quad T_2=40,000 \K  \\ % 3.45
   2.3 \times 10^{48}    & \quad  {\rm Hydrogen} \quad T_2=36,000 \K  \\ % 2.27
   8.2 \times 10^{47}    & \quad  {\rm Helium}   \quad T_2=40,000 \K  \\ % 8.18
   1.9 \times 10^{47}    & \quad  {\rm Helium}   \quad T_2=36,000 \K  \\ % 1.87
\end{array}\right.
\label{ni2}
\end{equation}

% ==========================================================
\section{THE IONIZING RADIATION PROPAGATION THROUGH THE WIND}
\label{ion}
% ==========================================================
\subsection{The Undisturbed Primary Wind}
%==========================================
Let us consider the distance to which the secondary ionizes the undisturbed
primary wind along a direction perpendicular to the orbital plane.
By undisturbed we mean that the influence of the secondary gravity and wind
on the primary wind is neglected.
The density of the secondary wind is very low, and it can be neglected
when calculating the recombination rate.
The density of the undisturbed primary wind as function of distance $y$
perpendicular to the orbital plane measured from the secondary is
$\rho_1(y)={\dot M_1}/ 4 \pi (r^2+y^2) v_1$,
where $r$ is the orbital separation.
The total hydrogen recombination rate per steradian along that direction is
\begin{equation}
\dot R_{\rm pole} =\alpha_B \int_0^\infty n_e n_H y^2 dy
= 0.22  \alpha_B  \left( \frac{\dot M_1}{4 \pi v_1 \mu m_H} \right)^2
 \int_0^\infty \frac{y^2}{(r^2+y^2)^{2}} dy,
\label{rh1}
\end{equation}
where $\alpha_B$ is the recombination coefficient and
$\mu m_H$ is the mean mass per particle in a fully ionized gas.
A similar expression can be derived for the recombination rate of $He^+$, assuming
that its abundance by number is $10 \%$.
Preforming the integral and substituting typical values gives for the recombination
rate per steradian
\begin{equation}
\dot R_{\rm pole} =
\left( \frac {\dot M_1}  {3 \times 10^{-4} M_\odot \yr^{-1}}  \right)^2
\left( \frac {v_1}  {500}  \right)^{-2}
\left( \frac {r}  {5 \AU}  \right)^{-1}
\s^{-1} {\rm sr}^{-1}\left\{
\begin{array}{ll}
   5.4 \times 10^{47} & \quad  {\rm Hydrogen}  \\  % 5.1
   5.7 \times 10^{46} & \quad  {\rm Helium}.  \\
\end{array}\right.
\label{rec2}
\end{equation}
The orbital separation of $5 \AU$ is the one at $\sim 35$~day before
and after periastron.
In deriving equation (\ref{rec2}) we assumed a spherical mass loss geometry and
ignored the dependence of the primary wind's density and velocity on latitude
(Smith et al. 2003).

\begin{figure}
\resizebox{0.82\textwidth}{!}{\includegraphics{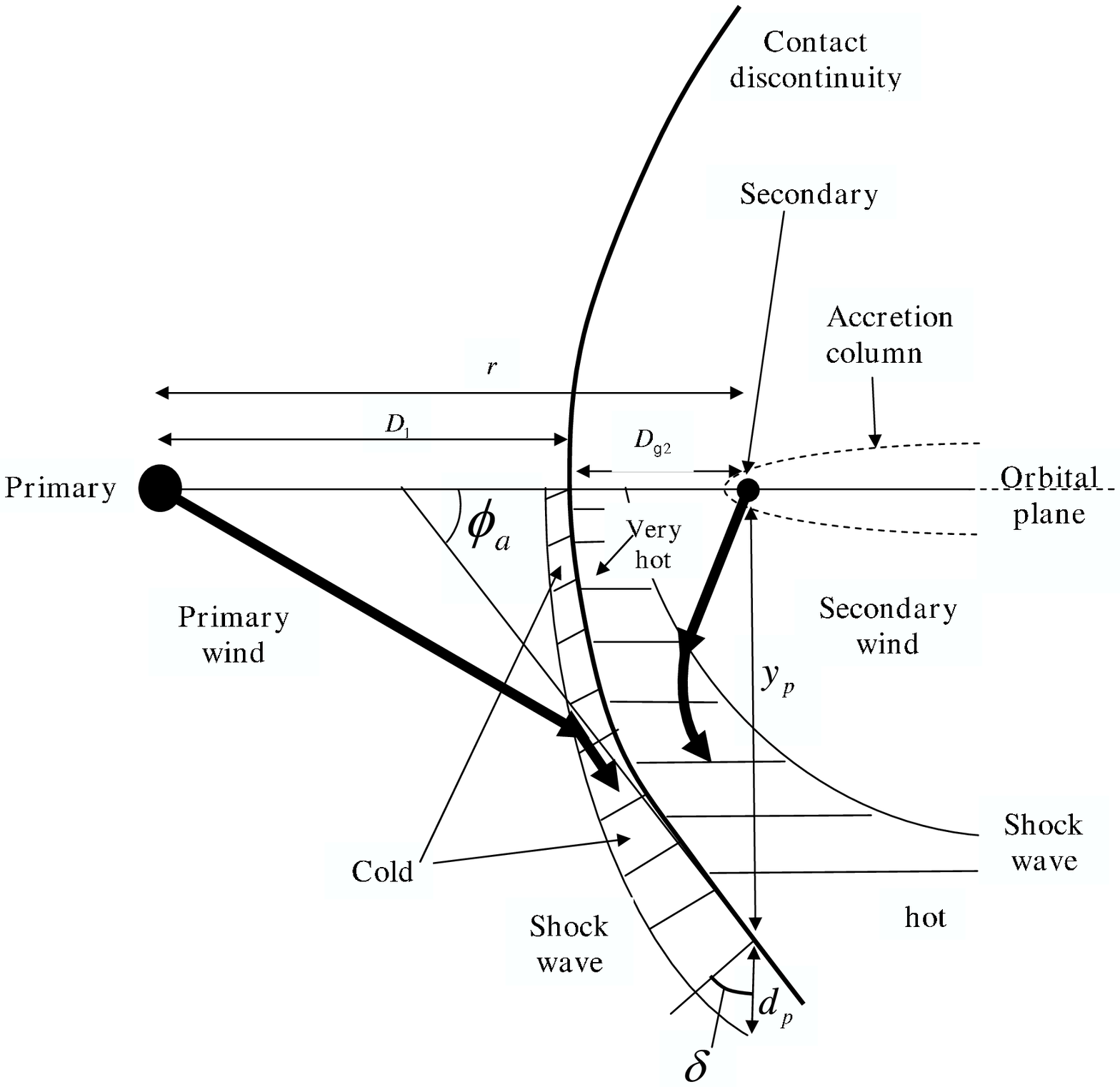}}
\vskip -2.2 cm
\caption{Schematic drawing of the collision region of the two stellar
winds and definition of several quantities. The two thick lines
represent winds' stream lines. The two shock waves are drawn only
in the lower half. The post-shock regions of the two winds are
hatched.
The dashed line shows the accretion column which exists,
according to the proposed model, only for $\sim 70-80$~days
during the accretion period which corresponds to the X-ray  minimum
and the spectroscopic event
(adopted from Akashi et al. 2006).
} \label{fwinds}
\end{figure}

Along a direction from the secondary to the opposite side of the primary
(right side in Figure \ref{fwinds}), the recombination rate through the
undisturbed primary wind, $\dot R_b$, is obtained by replacing $(r^2+y^2)^{2}$
with $(r+y)^{4}$ in the denominator of equation (\ref{rh1});
the recombination rate is smaller by a factor of $3 \pi /4$
\begin{equation}
\left( \frac{\dot R_b}{\dot R_{\rm pole}} \right)_0 = \frac{4}{3 \pi}=0.42,
\label{rr}
\end{equation}
where subscript zero indicates that this expression holds for the undisturbed primary wind.

Comparing equations (\ref{rec2}) and (\ref{rr}) with equation (\ref{ni2}) teaches us
that the secondary can (for the chosen parameters) ionize the undisturbed
primary wind to large distances along the polar directions and the
equatorial directions away from the primary.
The recombination rate is quite sensitive to the primary wind's properties.
An increase by a factor of $\sim 3$ in the ratio $\dot M_1/v_1$
(namely, an increase in the mass loss rate and/or a decrease in the wind speed)
will result in an ionization distance through the undisturbed primary wind smaller
than the distance to the regions that emit the emission lines, e.g., the Weigelt blobs,
when $r= 5\AU$.
For a mass loss rate as high as $\dot M_1 \ga 3\times 10^{-3} M_\odot \yr^{-1}$ the Str\"omgren radius
is within the wind at all orbital separations even for the undisturbed primary wind.
Martin et al. (2006c) suggested that at the beginning of the 20th century
the primary stellar mass loss rate was much higher, $\dot M_1 \sim 10^{-2} M_\odot \yr^{-1}$,
such that many emission lines were not observed at all.
We instead suggest that the secondary was cooler, hence had weaker ionizing
radiation (section \ref{summary}).
%====================================================================
%% \clearpage
%====================================================================

%=====================================
\subsection{The Shocked Primary Wind}
%=====================================
The secondary wind `cleans' the area behind the secondary (right side in Figure \ref{fwinds})
and compresses the primary wind along the contact discontinuity, increasing the recombination
rate there.
Therefore, the ratio between the recombination rate in the equatorial plane in the
direction away from the primary, $\dot R_b$, and that in the polar directions,
$R_{\rm pole}$, is much smaller than the value given by equation (\ref{rr})
for the undisturbed primary wind.
The half opening angle of the wind-collision cone (see Figure \ref{fwinds})
is $\phi_a \simeq 60 ^\circ$ (Akashi et al. 2006).
This implies that even as the secondary approaches or recedes periastron
the secondary fast wind will clean a large solid angle for ionization to propagate
almost unattenuated at low latitudes, unless the fast wind is shut down as assumed here
(see section 3.3).

The post-shock primary wind has a much higher density and hence its high recombination
rate must be considered.
The postshock region is unstable, and has a corrugated structure
(Pittard \& Corcoran 2002; Pittard et al. 1998).
For that, our calculation here is a crude estimate, but still teaches us
on the importance of the shock wave.
The post-shocked primary wind flows in a thin shell along the contact discontinuity.
Let the width of the conical shell along the polar direction at distance $y_p$
from the secondary be $d_p$; these quantities are defined in Figure \ref{fwinds}.
The primary wind rapidly cools to a temperature of $T_p \simeq 10^4 \K$,
and it is compressed by the ram pressure of the slow wind to a density $\rho_p$.
Neglecting first the magnetic pressure in the post-shock region,
$\rho_p$ is given by equating the thermal pressure $kT_p \rho_p/\mu m_H$ of the post-shock material
with the ram pressure of the primary wind $\rho_1 (v_{\rm wind1} \sin \beta)^2$,
where $\beta$ is the angle between the slow wind speed and the shock front at $y_p$,
and $v_{\rm wind1}$ is the pre-shock speed of the primary wind relative to the stagnation
point.
For the present purpose we can take $v_{\rm wind1} \simeq v_1$. From Figure 2 of
Pittard \& Corcoran (2002) we find $\beta (y_p) \sim 30 ^\circ$ and $y_p \sim 0.5-0.6 r$.
For a post-shock temperature of $T_p \simeq 10^4 \K$ we find
\begin{equation}
\left( \frac{\rho_p}{\rho_1} \right)_{\rm thermal} \simeq 470    %469
\left( \frac{v_1}{500 \km \s^{-1}} \right)^{2}
\left( \frac{\sin \beta}{0.5} \right)^{2}.
\label{shock1}
\end{equation}
Such a high density contrast is seen in Figure 2 of Pittard \& Corcoran (2002).
However, because of magnetic fields we don't expect such a large compression
behind the shock.
Typical pre-shock magnetic pressure to ram pressure ratio can be
$\eta_B \equiv P_{B0}/\rho_1 v_1^2 \sim 0.001-0.1$
(e.g., Eichler \& Usov 1993; Pittard \& Dougherty 2006).
The magnetic field component parallel to the shock is increased as it
is compressed when the density is increased in the shock wave.
For a random field we can take this component to contribute $1/3$ to the pre-shock pressure.
Equating the post-shock magnetic pressure to the wind's ram pressure we find the limit
on the compression factor imposed by the magnetic field to be
\begin{equation}
\left( \frac{\rho_p}{\rho_1} \right)_B \simeq \left( \frac{3}{\eta_B} \right)^{1/2}
\simeq 30 \left( \frac{\eta_B}{0.003} \right)^{-1/2}.
\label{shockB}
\end{equation}
More than that, the strong magnetic field can smooth the strong corrugated structure seen in
the simulations presented by Pittard \& Corcoran (2002).
The presence of the magnetic field is another source of the large uncertainties
involved in our calculation, and it can introduce large stochastic variations
on short time scale and from cycle to cycle.
Considering equations (\ref{shock1}) and (\ref{shockB}) we will use the scaling
${\rho_p}/{\rho_1}=100$.

If the post-shock gas outflows at a speed $v_d$ along the contact discontinuity,  then
mass conservation for the mass entering the region $y \le y_p$ reads
\begin{equation}
\dot M_1 (y_p) \equiv \frac{1}{2} \left[ 1- \frac{r}{ \left(r^2+y_p^2 \right)^{1/2}} \right]
\dot M_1  \simeq 2 \pi v_d \rho_p y_p d_p \cos \delta ,
\label{shock2}
\end{equation}
where the angle $\delta$ is defined in Figure \ref{fwinds}.
The velocity $v_d$ of the post-shock primary wind parallel to the shock front
is zero at the stagnation point (where the two winds momenta balance each other
along the symmetry axis), and increases to $\sim v_1$ at infinity.
Mixing between the two winds, as a result of instability  (the corrugated structure;
Pittard \& Corcoran 2002), will further accelerate the post-shock primary wind
(Girard \& Willson 1987).
This happens because the post-shock secondary wind expands faster than the
primary wind and because of its long cooling time it has a pressure gradient
parallel to the contact discontinuity.
We therefore can take $0.5 v_1/v_d \cos \delta \sim 1$.
Substituting typical values in equation (\ref{shock2}) gives
\begin{equation}
\frac {d_p}{y_p} \simeq
\frac{v_1}{v_d} \frac{0.5}{\cos \delta}
\frac{\rho_1}{\rho_p} \simeq \frac{\rho_1}{\rho_p}.
\label{shock3}
\end{equation}

The recombination rate per steradian for the undisturbed wind up to distance
$y=y_p$ from the secondary goes as
$\dot R_{0}(y_p) \simeq \alpha_B n_e n_p y_p^3 /3 \propto y_p^3 \rho_1^2/3$,
while in the post-shock shell at $y=y_p$ it is
$\dot R_{\rm shock-pole}(y_p) \simeq \alpha_B n_e n_p y_p^2 d_p  \propto y_p^2 d_p \rho_p^2$.
Using equation (\ref{shock3}) we find
$\dot R_{\rm shock-pole}(y_p) \simeq 3 (\rho_p/\rho_1) \dot R_{0}(y_p) $.
We find that the recombination rate in the postshock region is much higher
than that in the undisturbed slow wind occupying the same region.
Substituting typical values that we used above, with $y_p=0.55r$,
we find the recombination rate per steradian of
the shocked gas
\begin{equation}
\dot R_{\rm shock-pole}(y_p) \simeq
\left( \frac {\dot M_1}  {3 \times 10^{-4} M_\odot \yr^{-1}}  \right)^2
\left( \frac {v_1}  {500}  \right)^{-2}
\left( \frac {r}  {5 \AU}  \right)^{-1}
\frac{\rho_p}{100 \rho_1}
\s^{-1} {\rm sr}^{-1}\left\{
\begin{array}{ll}
    10^{49} & ~  {\rm Hydrogen}  \\  % 1.1
    10^{48} & ~ {\rm Helium}.  \\
\end{array}\right.
\label{rec3}
\end{equation}

Comparing equation (\ref{rec3}) with equation (\ref{ni2}) shows that for an orbital separation
of $r \simeq 5 \AU$ the recombination rate along the polar directions is about equal
to the ionization rate.
This implies that the ionization structure of the primary wind will change as
the two stars orbit each other, and that this structure is sensitive
to the primary wind properties: mass loss rate, speed, magnetic field.
This sensitivity can result in stochastic variation from cycle to cycle
and within one cycle.
The primary wind parameters, and in particular the compression ratio
within the shock can be constraint from the free-free radio emission
of $\eta$ Car. This will be the subject of a forthcoming paper.

%===========================================
\subsection{The Path Opened by the Secondary Wind}
%==============================================

\begin{figure}
\resizebox{0.68\textwidth}{!}{\includegraphics{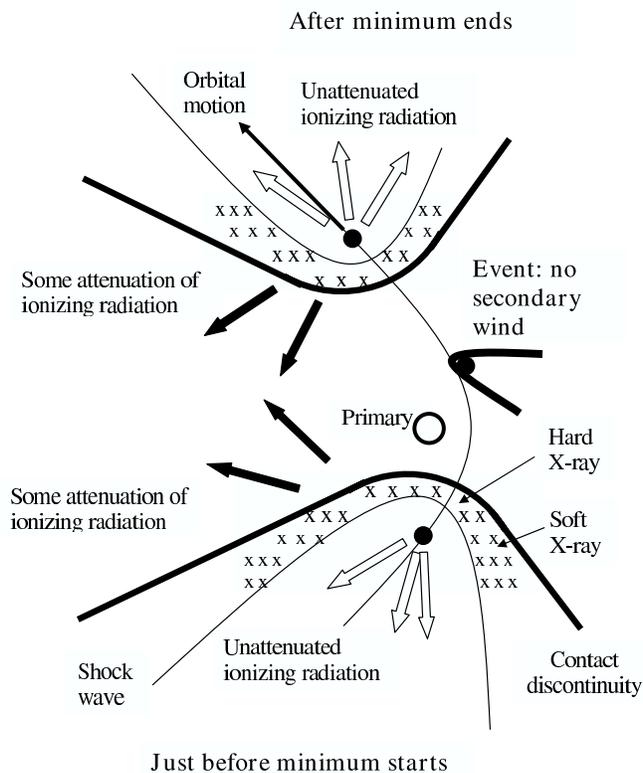}}
\vskip -2.2 cm
%{\includegraphics{etahef2.eps}}
\caption{Schematic drawing (not for scaling and not the exact
shock waves and contact discontinuity structures) of the flow
structure at three epochs: Just before and after the spectroscopic event,
or the X-ray minimum, when the two winds exist, and during the X-ray minimum,
when the secondary wind is assumed to be extinct. Note that
according to our model the X-ray minimum and the spectroscopic event
are not symmetric about periastron. The shocked primary wind is marked by
the thick arcs;
the X-ray emitting shocked secondary wind is in the region marked
by `x's'; the open and filled circles mark the positions of the
primary and secondary, respectively.
Secondary stellar ionizing radiation escaping through the secondary wind
suffers almost no attenuation (empty arrows).
During the event, the accretion column absorbs the radiation  along
that direction, while the dense primary wind absorbed the radiation
in the primary direction.
Hence, no hard ionizing radiation escape in the equatorial plane during
the event (adopted from Akashi et al. 2006).}
\label{fabsorb}
\end{figure}

The derivation of subsections (3.1) and (3.2)
is true for directions through the primary wind.
The secondary wind opens a large solid angle through which the secondary
ionizing radiation can freely expand to infinity.
For example, the radiation can reach the Weigelt blobs from where
high-ionization emission lines are observed (Hamann et al. 2005;
Hartman et al. 2005;  Johansson et al. 2006).
This is shown before and after periastron passage in Figure \ref{fabsorb}.
This solid angle is to a side opposing the primary wind
(the right side of the secondary star in Figure \ref{fwinds}),
and the asymptotic opening angle of this region is
$\phi_a \simeq 60^\circ$ (Akashi et al. 2006).
Therefore, if the secondary continues to blow its wind during the
spectroscopic event a large fraction of the ionizing radiation will reach
far regions in the equatorial plane even when the binary system is near periastron,
and there will be no large fading in high ionization emission lines.
The decrease in orbital separation can account for a slower variation,
like the slow decline in the $H_\alpha$ which starts 3 months before
the event (Davidson et al. 2005).
Three months before periastron passage the orbital separation is
$\sim 10 \AU$, and a reduction in the ionizing photons reaching large
distance can be significant (depending on the exact primary wind's
parameters).
%
%====================================================================
%====================================================================

{{{  The general flow structure and ionization and recombination rates
we deal with are far too sensitive to the stellar and wind parameters to be
modelled by a simple approach.
However, some estimates can be done. }}}
The primary is too cool to ionize helium at large distances {{{  and in all directions }}},
so the much hotter secondary must be the source of the harder ionizing radiation
(e.g., Steiner \& Damineli 2004; Verner et al. 2005).
{{{  Davidson \& Smith (2006) argued for a hot primary's equatorial photosphere,
$>20,000 \K$. }}}
In the past it was argued that most of the the photons with energy
$h \nu > 13.6 \ev$ emitted by the primary are absorbed by the wind
(Davidson \& Humphreys 1997; Hillier at al. 2001).
Hillier et al. (2001) found that the hydrogen in the wind becomes neutral at a
distance of $\sim 200 \AU$ from the star. They used a high mass loss rate
of $\dot M_1 = 10^{-3} M_\odot \yr^{-1}$, and small filling factor for the wind
(large clumping), and may have overestimated the global recombination rate in the wind.
Smith et al. (2003) found the stellar wind in the polar directions to be
denser than the wind blown at low latitudes and the equatorial direction.
The strong absorption seen in Balmer lines toward high latitudes indicate that
there is a large fraction of neutral hydrogen there.
Smith et al. (2003) argued that in the equatorial plane the wind might be largely
ionized, hence allowing the primary radiation to reach the Wigelt blobs.
The decrease in many emission lines with no decrease in the {{{  visual }}} continuum
(Martin \& Koppelman 2004) suggests that the cause of the spectroscopic
event is in the secondary star, and not in the much brighter primary star.

Najarro et al. (1997) studied the ionization structure of
the wind from P Cygni and found that the wind ionization structure as a result
of the ionizing radiation of the star blowing the wind is both
complicated and sensitive to the stellar parameters.
Although, our simple approach cannot be used to deduce the
exact ionizing effect of the primary radiation, as discussed above,
{{{  we can do a simple estimate.
For the parameters used here the number of hydrogen-ionizing photons emitted by the
secondary is $\sim 3$ times larger than that from the primary star
(say for $R_1=120 R_\odot$ and $T_{\rm eff1}=25,000 \K$).
The recombination rate along a line from the star is proportional to $1/r_{\rm min}$,
where $r_{\rm min}$ is the radius where the absorption stars. The primary
radius is $<0.6 \AU$, and therefore the recombination rate from the primary
to large distances is $\sim 10$ times larger than that from the secondary when the
secondary is at an orbital separation of $5 \AU$, the distance that was
used in sections (3.1) and (3.2). Because of the
primary wind-acceleration zone, where wind density is higher, the recombination
rate along a ray from the primary will be larger even.
With the primary ionization rate lower by a factor of  $\sim 3$ and recombination
rate along a ray larger by a factor of $>10$, compared with the secondary,
and using numbers we derived in equations (\ref{ni2}), (\ref{rec2}) and (\ref{rec3}),
we conclude that ionization by the primary is less important than that by the
secondary star.  }}}
Considering all these and the results of Verner et al. (2005), we concentrate on
ionization by the secondary star and the influence of the accretion flow near
periastron passages on the ionization process.

% ==========================================================
\section{THE IONIZATION DURING THE ACCRETION PHASE}
\label{acc}
% ==========================================================

There are several arguments why the spectroscopic event cannot by a simple
eclipsing event or a burial of the secondary inside the primary wind
(Stahl et al. 2005).
For example, in all directions the spectroscopic event has a fast onset
and a slow recovery (Stahl et al. 2005).
Such a behavior, as well as changes in emission lines intensities on time scales
shorter than the time scale over which the orbital separation changes, cannot
be accounted for by a simple orbital motion in an eccentric orbit.
For example, a decrease by a factor of two in the intensity of some lines
excited by the Lyman continuum occurs in $\sim 3$~day  (Hartman et al. 2005).
This cannot be attribute to the orbital motion alone, as
the fastest decrease in orbital separation occurs near periastron,
and a decrease by a factor of two, for example, in orbital separation
requires 18 days.
{{{  We note that this behavior does not contradict the shell ejection scenario
(Zanella et al. 1984).
The most severe problem for the shell ejection scenario is the X-ray light curve
(see section 1).}}}
The decrease in orbital separation can account for slower variations
as mention in Sec. \ref{ion}.
In addition, the spectroscopic event and the X-ray minimum last $\sim 70$~day.
Even if the event is symmetric around periastron (which is probably not),
the orbital separation $35~$day before or after periastron is $\sim 5 \AU$,
much larger than the primary radius, ruling out that the secondary is inside
the primary extended atmosphere.

Instead, a change in the gas flow must occur {{{  (Zanella et al. 1984; }}}
Stahl et al. 2005; Weis et al. 2005; Davidson et al. 2005).
Namely, a change in the velocity, mass loss rate, geometry of one or two of the
winds blown by the two stars, and/or a change in the interaction of the two winds.
We suggest that the main changes during the spectroscopic event can be attribute
to an accretion event which shuts down the secondary wind (Soker 2005b).
Figure \ref{fabsorb} schematically shows the evolution around the event.
According to the model (Soker 2005b; Akashi et al. 2006) during the spectroscopic
event the secondary does not blow its wind, but rather accretes mass, mainly from the
direction of the accretion column (the thick line attached to the
secondary during the event in Figure \ref{fabsorb};
the dashed line in Figure \ref{fwinds}).

The accretion process has two effects on the ionizing radiation emitted by the
secondary.
\newline
(1) {\it Absorption. }
After the secondary wind is shut down, according to the assumption
of the accretion model, and the primary wind material is accreted
from the up wind direction (primary direction) and from the back
through the accretion column, the large opening angle for the
ionizing radiation does not exit any more.
The density along the accretion column is very large
(e.g., Ishii et al. 1993; Ruffert 1996, 1999)
and no ionizing radiation will escape from that direction.
Now the absorption of ionizing radiation is much larger in the equatorial
plane and mid-latitude directions than along the polar directions.
First, the dense wind near the primary blocks the secondary radiation
passing too close to the primary.
Second, the dense accretion column is in the equatorial plane.
Third, as the secondary moves along its orbit the accretion column is
dragged behind, and covers other directions in the equatorial plane
and mid latitude (Mastrodemos \& Morris 1998, 1999).
\newline
(2) {\it Inflated envelope.}
The accreted matter has a non-negligible angular momentum.
Although not enough to form an accretion disk (Akashi et al. 2006),
it can still influence the accretion process.
The matter will concentrate in the equatorial plane, and will take some time
to reach equilibrium in the secondary's envelope, probably several time
the Keplerian orbital time, $\tau_{\rm Kep-2}=1.9~$day.
In addition, the secondary has a high radiation pressure on its surface,
as manifest in its strong wind. This might also lengthen the relaxation time of
the accreted mass onto the envelope.

During the event the Bondi-Hoyle mass accretion rate of the primary wind gas onto
the secondary changes from $\sim 0.2 \times 10^{-6} M_\odot \yr^{-1}$ to
$\sim 1.5 \times 10^{-6} M_\odot \yr^{-1}$ and then back to
$\sim 0.2 \times 10^{-6} M_\odot \yr^{-1}$ (Akashi et al. 2006).
At the onset of accretion, as the winds collision region collapses onto the
secondary, the accretion rate is higher than the Bondi-Hoyle rate.
We therefore scale accretion rate with
$\dot M_{\rm acc}= 10^{-6} M_\odot \yr^{-1}$.
We assume that this material reaches the secondary at a high speed,
and encounters a shock wave.
If this gas reaches the secondary at the free fall velocity of
$v_{ff2}=760 \km \s^{-1}$ (for a secondary mass of $M_2=30 M_\odot$ and a
radius of $R_2=20 R_\odot$), its radiative cooling time after the shock
would be $T_{\rm cool} \sim 2000 \s$. The distance the flow traverse during that
time is $\sim t_{\rm cool} v_{ff2} \simeq 2 R_\odot$.
Radiative breaking (Gayley et al. 1997), i.e., the radiation pressure
of the secondary, will slow down the accreted mass and will increase the density
calculated here.

We suggest that the accreted mass forms a blanket at a distance of
a few solar radii, or $\sim 0.1-0.3 R_2$, around
the secondary, and takes a few dynamical time scales to relax,
$\tau_{\rm relax} \sim 5~$day; the total mass in the blanket is
$\tau_{\rm relax} \dot M_{\rm acc}$.
The optical depth of this blanket, for an opacity of $\kappa=0.4$, is
\begin{equation}
\tau = 0.45
\left( \frac {\dot M_{\rm acc}}  {10^{-6} M_\odot \yr^{-1}}  \right)
\left( \frac {t_{\rm relax}}  {5~{\rm day}}  \right)
\left( \frac {R_2}  {20 R_\odot}  \right)^{-2}.
\label{opac}
\end{equation}
The density of this material is $\sim 10^{-11} \g \cm^{-1}$, compared with the
secondary's photospheric density of $\rho_{\rm photosphere} \simeq 2 \times 10^{-10} \g \cm^{-1}$.

The conclusion is that during the accretion phase the effective photosphere of the secondary
star is at $R_{p2} \sim 1.1-1.3 R_2$, and its effective temperature
is somewhat lower $T_{\rm eff2} \simeq 0.85-0.95 T_2$.
This can substantially reduce the energetic photon flux at
$h \nu > 24.6 \ev$.
For example, using equation (\ref{nerg}) we find that in cooling from
$T_2=40,000 \K $ to $T_{\rm eff2} =37,000~(36,000) \K$, the ionizing flux from the
secondary decreases to $0.74$ ($0.66$) times its initial value for $h \nu > 13.6 \ev$,
and to $0.42$ ($0.23$) for $h \nu > 24.6 \ev$.
If the photospheric cooling is from $T_2=38,00 \K $ to $T_{\rm eff2} =35,500 \K$,
the ionizing flux from the secondary decreases to $0.74$ times its
initial value for $h \nu > 13.6 \ev$, and to $\sim 0.22$ for
$h \nu > 24.6 \ev$.
The effect is stronger if we take $T_2=37,200 \K$ as suggested by Verner et al.
(2005).
It is clear that the decrease in the flux of energetic photons, $h \nu > 24.6 \ev$,
is much larger that that of lower energy photons.

Both effects, absorption and inflating the photosphere, cause more attenuation of the
ionizing radiation in the equatorial plane and mid latitudes than in the polar directions.
In addition, before the accretion starts there is a large opening angle
formed by the low density secondary wind, through which ionizing radiation reaches
large distances unattenuated.
This opening solid angle is in the equatorial plane to mid-latitudes
(right side in Figure \ref{fwinds}; empty double arrows in Figure \ref{fabsorb}).
Therefore, our model accounts for the observation that the spectroscopic event is
more pronounced in the equatorial and mid latitudes directions
than in the polar directions (Stahl et al. 2005; Weis et al. 2005).
This is supported by the large changes observed in the Weigelt blobs
(e.g., Hamann et al. 2005) which are thought to reside in the equatorial plane
(Davidson et al. 1997).
Note that if accretion does not occur during the spectroscopic event, then
according to equations (\ref{rr}) and (\ref{rec3}) the attenuation along the polar directions is
expected to be larger than that in the back direction in the equatorial plane.
This will result in larger variations along the polar directions,
contrary to observations.

%%%  The accretion rate is mach too low to cause the star to expand
%%% (Flannery \& Ulrich 1977).
%%%  \reference{}  Flannery, B. P. \& Ulrich, R. K.1977, ApJ, 212, 533
%%%

% ==========================================================
\section{SUMMARY}
\label{summary}
% ==========================================================

Our goal was to explain the rapid and large decrease in the intensity of high-ionization
emission lines starting with the spectroscopic event.
We were not aiming at explaining the relative intensities of different lines
or their exact temporal behavior, but rather to show that the accretion model
can account for the basic behavior of the fading high-ionization lines.
The fading of these high-ionization emission lines serves as the definition
of the spectroscopic event.
We appeal to the accretion model for the spectroscopic event
(Soker 2005b; Akashi et al. 2006; Soker \& Behar 2006).
In that model some of the characteristics of $\eta$ Car during the spectroscopic
event are attributed to an accretion process where the secondary star is accreting
mass from the primary wind instead of blowing its fast wind
(Soker 2005b; Akashi et al. 2006; Soker \& Behar 2006).

In the present papers we studied the basic behavior of the ionizing radiation
emitted by the secondary star (the companion).
We found that the accretion event has two effects that substantially reduce the
high-energy ($h\nu > 24.6 \ev$) ionizing radiation flux that escape the vicinity
of the binary system.
The same effects apply, but to lesser degree, to the lower energy photons of
$24.6 \ev > h\nu >13.6 \ev$.
\begin{enumerate}
\item {\it Denser circumbinary material.} Without accretion the secondary wind
flows away from the binary system (to the right side in Figure \ref{fwinds})
and `cleans' a path for the secondary's ionizing radiation to reach large distances
(depicted by the empty arrows in Figure \ref{fabsorb}).
In that case the attenuation expected along the equatorial plane is less than
that along the polar directions (equations \ref{rr} and \ref{rec3}),
leading to larger variations in the intensities of the high-ionization
emission lines in the polar directions as the secondary orbit the primary.
Observations show that the spectroscopic event is less pronounced in the polar
directions than in the equatorial to mid latitude directions
{{{  (Smith et al. 2003; }}} Stahl et al. 2005; Weis et al. 2005).
Therefore, the expected behavior of this model without accretion
is contrary to observations.
When accretion is included, larger attenuation is expected, mainly in the
equatorial plane to mid latitude directions.
The reason is that the accreted mass forms a dense region behind the secondary,
the accretion column (dashed line in Figure \ref{fwinds}), that has a high
recombination rate, and hence prevents a large fraction of the high-energy
ionizing radiation to reach large distances in and near the equatorial plane.
\item  {\it Inflated envelope.}
The accreted matter has non-negligible angular momentum (Akashi et al. 2006).
The matter will concentrate in the equatorial plane, and will take some time
to reach equilibrium in the secondary's envelope, probably several times
the Keplerian orbital time.
In addition, the secondary has a high radiation pressure on its surface,
as manifest in its strong wind.
We propose that instead of an acceleration zone of the secondary wind,
a blanket is formed during the accretion process, that moves the
effective secondary's photosphere to be further out.
This reduces the effective temperature and the flux of high-energy photons,
mainly in the equatorial plane.
\end{enumerate}

These effects influence the energetic ionizing radiation, and as a result
of that the behavior of high-ionization emission lines in a way
compatible with observations:
($i$) Fading of high ionization lines at the same time the X-ray is
declining to its long minimum; ($ii$) The fading of these lines
is less pronounced in the polar directions.

The orbital motion brings the secondary into denser regions of the primary
wind as it approaches periastron, and can explain relatively slow variations
over the 5.54 orbital period.
However, the variations due to orbital motion alone cannot account for
the sharp disappearance of high-ionization emission lines during the
spectroscopic event.

We emphasize that we were not aiming at explaining all complications involved in
the variations of lines intensities.
In particular we did not consider the following processes.
\newline
(1) {\it Evolution of the primary star.} The secular variations
related to the evolution of the primary star and its wind as an
LBV star (Davidson et al. 2005; Martin \& Koppelman 2004; Martin
et al. 2006c). We do note, however, that the evolution of the
secondary can also affect the secular evolution of $\eta$ Car.
Most likely the secondary accreted several solar masses during the
Great Eruption of the 19th century (Soker 2001, 2007). The
accretion event $\sim 160$ years ago drove the secondary away from
equilibrium, and spun it up. It is possible that the secondary
swelled, and it is contracting back since then. The contraction
implies that the secondary photosphere has been heating up over
the last 150 years. From angular momentum conservation we expect
it to spin faster. This evolution implies that the high-energy
ionization radiation flux is increasing, and the fast rotation may
lead to hotter polar caps, and hence stronger ionizing radiation
along the polar directions.
{{{  The evolution of the secondary described above is in accord with the absence of
high ionization lines before 1941 (Feast et al. 2001; Humphreys \& Koppelman 2005).
The secondary was larger, hence cooler, and could not ionized the region now
responsible for the high-ionization emission lines.
The accretion events took place before 1941, but they could not be observed via
these emission lines. }}}

{{{  Other processes not considered in the paper include: }}}
\newline
(2) {\it Density inhomogeneities in the equatorial plane.}
Before the accretion phase starts radiation from the secondary is
less attenuated along direction through the secondary wind
(right side in Figure \ref{fwinds}1).
During the orbital motion this solid angle points to different directions in
the orbital plane and around it
{{{  Smith et al. (2004). }}}
Density inhomogeneities in the equatorial plane
will results in stochastic variations in line intensities.
\newline
(3) {\it Variable winds.} Likewise, stochastic variation in the primary wind can lead to
stochastic attenuation of the ionizing radiation.
\newline
(4) {\it Non-spherical winds.}
The dependence of the primary wind properties on latitude (Smith et al. 2003)
were ignored in our treatment.
{{{  In any case, Smith et al. (2003) find the latitude dependance of the primary wind
   properties to be large near apastron, whereas in the accretion the properties
   near periastron are important. Also, it is quite possible that this latitude
   dependance is influenced by the secondary star (Soker 2003). }}}
\newline
(5) {\it Collimated outflow from the secondary.}
Other types of outflows suspected from the secondary star (Behar et al. 2006),
that can change the `cleaned' solid angle for ionizing radiation also influence he line
intensities.

I thank Amit Kashi for helpful discussions,
{{{  and an anonymous referee for detailed and very helpful clarifications and
comments. }}}
This research was supported by a grant from the
Asher Space Research Institute at the Technion.

\end{document}